\documentstyle[preprint,prl,aps]{revtex}
 \begin{document}
 \draft
 
 \title{Characterization of a periodically
         driven chaotic dynamical system}
 
\author { A.~Crisanti$^1$, M.~Falcioni$^2$, G. Lacorata$^3$, R. Purini$^3$ \\
 and A.~Vulpiani$^1$}
 \date{May, 1996}
 
 \maketitle

\begin{tabular}{ll}
$^1$ & Dipartimento di Fisica, Universit\`a ``La Sapienza'',
          I-00185 Roma, Italy \\
     & and Istituto Nazionale di Fisica della Materia, Sezione di Roma \\
$^2$ &  Dipartimento di Fisica, Universit\`a ``La Sapienza'',
          I-00185 Roma, Italy \\
     & and Istituto Nazionale di Fisica Nucleare, Sezione di Roma \\
$^3$ & Istituto di Fisica dell' Atmosfera - CNR, P.le L. Sturzo 31
          I-00144 Roma, Italy
\end{tabular}

 \begin{abstract}
 We discuss how to characterize the behavior of a chaotic 
 dynamical system depending on a parameter that varies 
 periodically in time. In particular, we study the
 predictability time, the correlations 
 and the mean responses, by defining  a local--in--time 
 version of these quantities. In systems where the time 
 scale related to the time periodic variation of the 
 parameter is much larger than the ``internal'' time scale, 
 one has that the local quantities strongly depend on 
 the phase of the cycle. In this case, the standard 
 global quantities can give misleading information. 
 \end{abstract}
 
 
 \section{Introduction}
 The forecast of the behavior of a system when its evolution 
 law is known, is a problem with an obvious interest in many 
 fields of scientific research. Roughly speaking, within this 
 problem two main areas of investigation may be identified:

 \begin{description}

 \item{A)} The definition of the ``predictability time''. 
  If one knows the initial state of a system, with a precision 
 $\delta _o = |\delta \bbox{x} (0)|$, what is the maximum time 
 $T_p$ within which one is able to know the system future state 
 with a given tolerance $\delta _{{\rm max}}$ ? 

 \item{B)} The understanding of the relaxation properties.  What  
  is the relation between the mean response of a system to an 
  external perturbation and the features of its unperturbed 
  state \cite{Leith78}? Using the terminology of statistical mechanics, 
  one wants to reduce ``non equilibrium'' properties, such as relaxation 
  and responses, to  ``equilibrium'' ones, such as correlation 
  functions \cite{Kubo85}. 

 \end{description}

 A remarkable example of type-$A$ problem is the weather forecasting, 
 where one has to estimate the maximum time for which the prediction 
 is enough accurate. As an example of type-$B$ problem, of 
 geophysical interest, one can mention a volcanic eruption which 
 induces a practically instantaneous change in the temperature. 
 In this case it is relevant to understand how the difference 
 between the atmospheric state after the eruption and the 
 hypothetical unperturbed state without the eruption evolves 
 in time. In practice one wants to understand how a system 
 absorbs, on average, the perturbation $\delta f(\tau)$ of a 
 certain quantity $f$ -- {\it e.g.} the temperature -- just 
 looking at the statistical features, as correlations, of 
 $f$ in the unperturbed regime. This is the so called 
 fluctuation/relaxation problem \cite{Leith75}.
 
 As far as problem-$A$ is concerned, in the presence of 
 deterministic chaos, a rather common situation, the distance 
 between two initially close trajectories diverges exponentially: 
 \begin{equation}
 |\delta  \bbox{x} (t)| \sim \delta _o \exp (\lambda t) , 
 \label{eq:1}
 \end{equation}
 where $\lambda$ is the maximum Lyapunov 
 exponent of the system \cite{Bene76}. 
 From (\ref{eq:1}) it follows:
 \begin{equation}
 T_p \sim { 1 \over \lambda} 
 \ln \left( {\delta _{max} \over \delta _o }\right) . 
 \label{eq:2}
 \end{equation}
 Since the dependence on $\delta _{max}$ and $\delta _o$ 
 is very weak, $T_p$ appears to be proportional to the inverse of 
 the Lyapunov exponent. We stress however that (\ref{eq:2}) is just 
 a naive answer to the predictability problem, since it does not 
 take into account the following relevant features of the 
 chaotic systems: 
 \begin{description}
 \item {i)} The Lyapunov exponent is a global quantity, 
 {\it i.e.} it measures the average exponential rate of 
 divergence of nearby trajectories. In general there are 
 finite-time fluctuations of this rate, described by means 
 of the so called {\it effective} Lyapunov exponent
 $\gamma _t (\tau)$. This quantity depends on both the time 
 delay $\tau$ and the time $t$ at which the perturbation acted
 \cite{Pavu87}. Therefore the predictability time $T_p$ 
 fluctuates, following the $\gamma$-variations \cite{Cri93}. 
 \item {ii)} In systems with many degrees of freedom one has to 
 understand how a perturbation grows and propagates trough the 
 different degrees of freedom \cite{Pik92}. In fact, one can 
 be interested in the prediction on certain variables, {\it e.g.},  
 those associated  with large scales in weather forecasting,
 while the perturbations act on a different set of variables, 
 {\it e.g.}, those associated to small scales. 
 \item {iii)} If one is interested into non-infinitesimal 
 perturbations, and the system possess many characteristic times, 
 such as the eddy turn-over times in fully developed turbulence, 
 then $T_p$ is determined by the detailed mechanism of propagation 
 of the perturbations through different degrees of freedom, due to 
 nonlinear effects. In particular, $T_p$ may have no relation 
 with $\lambda$ \cite{Aubo96}. 
 \end{description}

 In addition to these points, when the evolution of a system 
 is ruled by a set of differential equations, 
 \begin{equation}
               d  \bbox{x}/dt=  \bbox{f} (\bbox{x},t) ,
 \label{eq:3}
 \end{equation}
 which depend periodically on time,  
 \begin{equation}
               \bbox{f} (\bbox{x},t+T) = \bbox{f} (\bbox{x},t) ,
 \label{eq:4}
 \end{equation}
 one can have a kind of ``seasonal'' effect, for which the 
 system shows an alternation, roughly periodic, of low and 
 high predictability. This happens, for example, in the 
 recently studied case of stochastic resonance in a chaotic 
 deterministic system, where one observes a roughly periodic 
 sequence of chaotic and regular evolution intervals \cite{Cri94}. 
 
 As far as problem-B is concerned, it is possible to show
 that in a chaotic system with an invariant measure $P(\bbox{x})$, 
 there exists a relation between the mean response 
 $\langle \delta x_j(\tau)\rangle _P$ after a time $\tau$ 
 from a perturbation $\delta x_i (0)$, and a suitable correlation 
 function \cite{Faca90}. Namely, one has the following equation: 
 \begin{equation}
  R_{i j} (\tau) \equiv { \langle \delta x_j (\tau) 
    \rangle _P  \over \delta x_i (0) } 
   =\left\langle x_j (\tau) {\partial S (\bbox{x} (0)) 
   \over \partial x_i}\right\rangle _P , 
 \label{eq:fluri}
 \end{equation}
 where $S (\bbox{x}) = -\ln P(\bbox{x}) .$
 Equation (\ref{eq:fluri}) ensures that the mean relaxation 
 of the perturbed system is equal to some correlation 
 of the unperturbed system. Of course since, in general, one 
 does not know $P(\bbox{x})$, eq. (\ref{eq:fluri}) provides 
 only a qualitative information. 
 
 In this paper we discuss how one has to reformulate the 
 predictability problem, both of type $A$ and of type $B$, 
 for systems where the ``seasonal'' effects are important, 
 {\it i.e.}, when some relevant characteristic times change, 
 in a substantial way, periodically  in time. 
 
 In section \ref{sec:frame} we discuss how to approach the 
 predictability problem in systems with time periodic evolution laws. 
 Section \ref{sec:toym} is devoted to the study of a toy model 
 whose behavior, in spite of the simplicity of the model, catches 
 the basic features of a system with seasonal effects. 
 In section \ref{sec:numsim}
 we show the results of numerical experiments, that illustrate the 
 relevance of the concepts introduced in section \ref{sec:frame}.

 \section{The characterization of systems with periodic effects}
 \label{sec:frame}

  We consider systems in which one can identify two time-scales: 
  the first, that we call $T_E$, can be thought as due to the 
  coupling with an external time-dependent driver; the second, 
  an ``internal'' time-scale $T_I$, characterizes the system
  in the limit of a constant external coupling, that we call 
  the ``stationary limit''. In the following, the external 
  time dependence will be assumed periodic with period $T$, 
  so that $T_E=T$.  
 
  If the system, in the stationary limit, is chaotic, 
  one can take as the internal time-scale the Lyapunov time, 
  {\it i.e.}, the inverse of the maximum Lyapunov exponent,
  $T_I\sim 1/\lambda$.

  In the case $T_E \gg T_I$ one can assume that the 
  system is adiabatically driven through different dynamical 
  regions, so that on observation times short with respect 
  to the long external time, it evolves on a local (in time) 
  attractor. If during its tour across the slowly changing 
  phase space the system visits regions where the effective 
  Lyapunov exponents are enough different, then the 
  system shows up sensibly different predictability times, 
  that may occur regularly when the driver is time-periodic.
  Consider, for instance, a slight modification of the 
  Lorenz model \cite{Lor63}, which is the first geophysical 
  dynamical system where deterministic chaos has been observed: 
\begin{equation}
  \left\{
         \begin{array}{ll}
              dx/dt=& 10 \, (y-x) \\
              dy/dt=& -x \, z + r(t) \, x - y \\
              dz/dt=& x \, y -{8\over 3} \, z
         \end{array}
  \right.
\label{eq:L3}
\end{equation}
where the control parameter  has a periodic time variation:
\begin{equation}
  r(t)=r_o -A \, \cos(2 \pi t/T).
\label{eq:L4}
\end{equation}
Since this model describes the convection of a fluid heated 
from below between two layers  whose temperature difference is 
proportional to the Rayleigh number $r$, the periodic 
variations of $r$ roughly  mimic the seasonal changing on the 
solar heat inputs. When $r_o$ is close to the threshold 
$r_{cr}=24.74$, where in the standard Lorenz model a transition 
takes place from stable fixed points to a chaotic attractor, 
and the amplitude $A$ of the periodic forcing  
is such  that $r(t)$ oscillates below and above $r_{cr}$, a good 
approximation of the  solution for very large $T$, may be given by 
\begin{equation}
  x(t)=y(t)=\pm \sqrt{ {8\over 3} (r(t)-1)} \qquad \qquad 
  z(t)=r(t)-1 ,
\label{eq:L5}
\end{equation}
which is obtained from the fixed points of the standard 
Lorenz model by replacing $r$ with $r(t)$. The stability 
of this solution is a rather complicated issue, which 
depends on the values of $r_o$, $A$ , and $T$. It is 
natural to expect that if $r_o$ is larger 
than $r_{cr}$ the solution is unstable. In this case, for $A$ 
large enough (at least $r_o-A < r_{cr}$) one has a mechanism 
similar to stochastic resonance in bistable systems with 
random forcing \cite{Cri94}. The value of $T$ is crucial: 
for large $T$ the systems behaves as follows. If 
\begin{equation}
    T_n \simeq nT/2 - T/4 \qquad (n=1,2,\dots)
\end{equation}
are the times at which $r(t)=r_{cr}$, one can naively expect 
that for $0<t<T_1$ -- when $r(t)$ is smaller than $r_{cr}$ -- 
the system is stable and the trajectory is close to one 
of the two solutions (\ref{eq:L5}), while for $ T_1<t<T_2$ 
-- when $r(t)>r_{cr}$ -- both solutions (\ref{eq:L5}) are 
unstable and the trajectory relaxes toward a sort of 
``adiabatic'' chaotic attractor. The chaotic attractor 
smoothly changes at varying of $r$ above the threshold 
$r_{cr}$, but if $T$ is large enough this dependence 
in first approximation can be neglected. When $r(t)$ becomes 
again smaller than $r_{cr}$, the ``adiabatic'' attractor 
disappears and, in general, the system is far from the stable 
solutions (\ref{eq:L5}); but it relaxes toward them, being 
attractive. If the half-period is much larger than the 
relaxation time, in general, the system follows one of the 
two regular solutions (\ref{eq:L5}) for $T_{2n+1} < t < T_{2n+2}$. 
However, there is a small but non-zero  probability that the system 
has no enough time to relax to (\ref{eq:L5}) and its evolution 
remains chaotic. Figure \ref{fig:1}  shows the time evolution of 
the variable $z$, for $r_o = 25.5$ and $A=4$, in the cases 
$T= 300$ (a) and $T= 1600$ (b). They provide an unambiguous 
numerical evidence that the jumps  from the  chaotic to the 
regular behavior (and the contrary) are well synchronized 
with $r(t)$, with probability close to $1$ when the forcing 
period $T$ is very long (see fig.~\ref{fig:1}~b). On the other hand, 
for small value of $T$ the system often does not perform the 
transition from the chaotic to the regular behavior, as 
seen in fig.~\ref{fig:1}~a.

It is worth stressing that  the system is chaotic. In both cases, 
in fact the first Lyapunov exponent is positive. 

  It is rather clear from this example that the 
  Lyapunov exponent is not able to characterize the above 
  behavior, since it just refers to a very long time 
  property of the system, {\it i.e.}, a property involving times 
  longer than $T$. A more useful and detailed information 
  can be obtained by computing a ``local'' average 
  of the exponential rate of divergence for initially close 
  trajectories. By this we mean an average which explicitly 
  depends on the time $t_o$, modulus the external period $T$, 
  to test the behavior of the system in the different states 
  of the external driver. In this way one can make 
  evident different behaviors, if any, of the system. 

 We therefore define the mean effective Lyapunov exponent, for the 
 time $t_o \in [0,T]$ and for a delay $\tau$, as
 \begin{equation}
 \langle \gamma(\tau)\rangle _{t_o} = \lim_{N \to \infty} 
 {1 \over N} \sum_{k=0}^{N-1}
  \gamma(t_o + k T, \tau),
 \label{eq:5}
 \end{equation}
 where
 \begin{equation}
 \gamma(t,\tau)={1 \over \tau} \ln {|\delta \bbox{x} (t+\tau)| \over
 | \delta \bbox{x} (t) | }, 
 \label{eq:6}
 \end{equation}
 is the local expansion rate, and $\delta {\bf x}(t)$ 
 evolves according to the linear equation
 \begin{equation}
 {d \over d t} \delta x _i (t) = \sum _j 
  {\partial f_i (\bbox{x}(t),t) \over  \partial x_j } \delta x_j. 
 \label{eq:}
 \end{equation}
 From this definition it is clear that 
 $\langle \gamma(\tau)\rangle _{t_o}$  measures the growth of 
 the distance between two trajectories that differ by 
 $| \delta \bbox{x} (t) |$ when the external driver passes through 
 a fixed value (or state). The maximum Lyapunov exponent of the 
 system gives the global average of $\langle \gamma(\tau)\rangle _{t_o}$: 
 \begin{equation}
 \lambda = \lim _{\tau \to \infty} \langle\gamma(\tau)\rangle _{t_o}
    ={1\over T}\int^T _0 \langle \gamma(\tau)\rangle _{t_o} d t_o .
 \label{eq:7}
 \end{equation}
 If one is interested on predictability for times
 much smaller then $T$, $\langle \gamma(\tau)\rangle _{t_o}$ 
 is a more appropriate quantity than $\lambda$, since it 
 distinguishes among different regimes. For example, 
 in the system (\ref{eq:L3}) discussed above, for the given 
 values of the parameters, one has $\lambda > 0$, but 
 $\langle \gamma(\tau)\rangle _{t_o} < 0$ when 
 $t_o \in [(n-1/4)T , (n+1/4)T]$. In the case of weather 
 forecasting, different values of $\gamma$ for different $t_o$, 
 correspond to different degree of predictability during the year. 

  As far as the response properties are concerned, 
  we expect that in a chaotic system the hypothesis of
  existence of ``adiabatic'' attractors implies that a  
  fluctuation/relaxation relation holds also as a time-local 
  property, provided one uses correlation and response 
  functions computed according a local, not a global, average. 
  So besides the usual correlation function between the 
  variables $x_i$ and $x_j$, 
 \begin{equation}
  C^{(G)} _{i j} (\tau) =
     \overline{ x_i (t) x_j (t+\tau)} -
         \overline{x_i} \, \overline{x_j}, 
 \label{eq:8}
 \end{equation}
 where $\overline{(\cdot)}$ indicates the global average,
 \begin{equation}
   \overline{A_i}= \lim _{t \to \infty} 
         {1\over t} \int^{t} _o 
         A_i (t')  d t' , 
 \label{eq:9}
 \end{equation}
 we introduce their correlation on a delay $\tau$ 
 after the time $t_o \in [0,T]$: 
 \begin{equation}
 C_{ij}(t_o,\tau)=\langle x_i(t_o)x_j(t_o+\tau)\rangle _{t_o} -
      \langle x_i \rangle _{t_o} \langle x_j\rangle _{t_o} ,
 \label{eq:10}
 \end{equation}
 where the local average is defined as in (\ref{eq:5})
 \begin{equation}
 \langle A \rangle _{t_o}=\lim_{N \to \infty}{1 \over N} 
  \sum_{k=0}^{N-1} A(t_o + k T). 
 \label{eq:11}
 \end{equation}
 In a similar way, one can consider two different kinds of 
 mean response function of the variable $x_i$ to a small 
 perturbation $\delta x_j$: the global average response, 
 \begin{equation}                                               
  R^{(G)} _{i j} (\tau) = {
     \overline{ \delta x_i (t+\tau) }\over \delta x_j (t) },
 \label{eq:12}
 \end{equation}
  and the local average response for the time $t_o$,
 \begin{equation}
  R_{i j} (t_o, \tau) = { \langle \delta x_i (t_o +\tau) \rangle _{t_o}
   \over \delta x_j (t_o) }. 
 \label{eq:13}
 \end{equation}
 The quantity (\ref{eq:12}) gives the mean response, after 
 a delay $\tau$, to a perturbation occurred at the time $t$, 
 chosen at random, {\it i.e.}, with uniform distribution 
 in $[0, T]$. We shall see that $R^{(G)} _{i j} (\tau)$ 
 can be rather different, even at a qualitative level, 
 from $R_{i j} (t_o, \tau) $. 

 \section{A toy model}
\label{sec:toym}

The ideas discussed in Sec. \ref{sec:frame} are illustrated 
here by means of a simple model. Consider the 
Langevin equation \cite{Wax} 
\begin{equation}
\label{eq:toy1}
\frac{d}{dt}\,q(t) = -a(t)\,q(t) + \xi(t) ,
\end{equation}
where $\xi(t)$ is $\delta$-correlated white noise, {\it i.e.},
$\xi(t)$ is a gaussian variable with
\begin{equation}
\label{eq:white}
   \langle\xi(t)\rangle = 0, \qquad 
   \langle\xi(t)\,\xi(t')\rangle = 2\Gamma\,\delta(t-t') ,
\end{equation}
and the coefficient $a(t)$ is a periodic 
function of period $T$: $a(t+T) = a(t).$ We require that 
\begin{equation}
\label{eq:toy3}
 \int_{0}^{T}dt\, a(t) > 0 ,
\end{equation}
to ensure a well defined asymptotic probability distribution for the
stochastic process given by eq. (\ref{eq:toy1}).
Moreover, we assume a slow variation of $a(t)$, {\it i.e.}
\begin{equation}
\label{eq:slow}
  \min_t\, a(t) \gg \frac{1}{T} , 
\end{equation}
so that, by making an adiabatic approximation, a 
``local'' probability distribution exists at any time.      

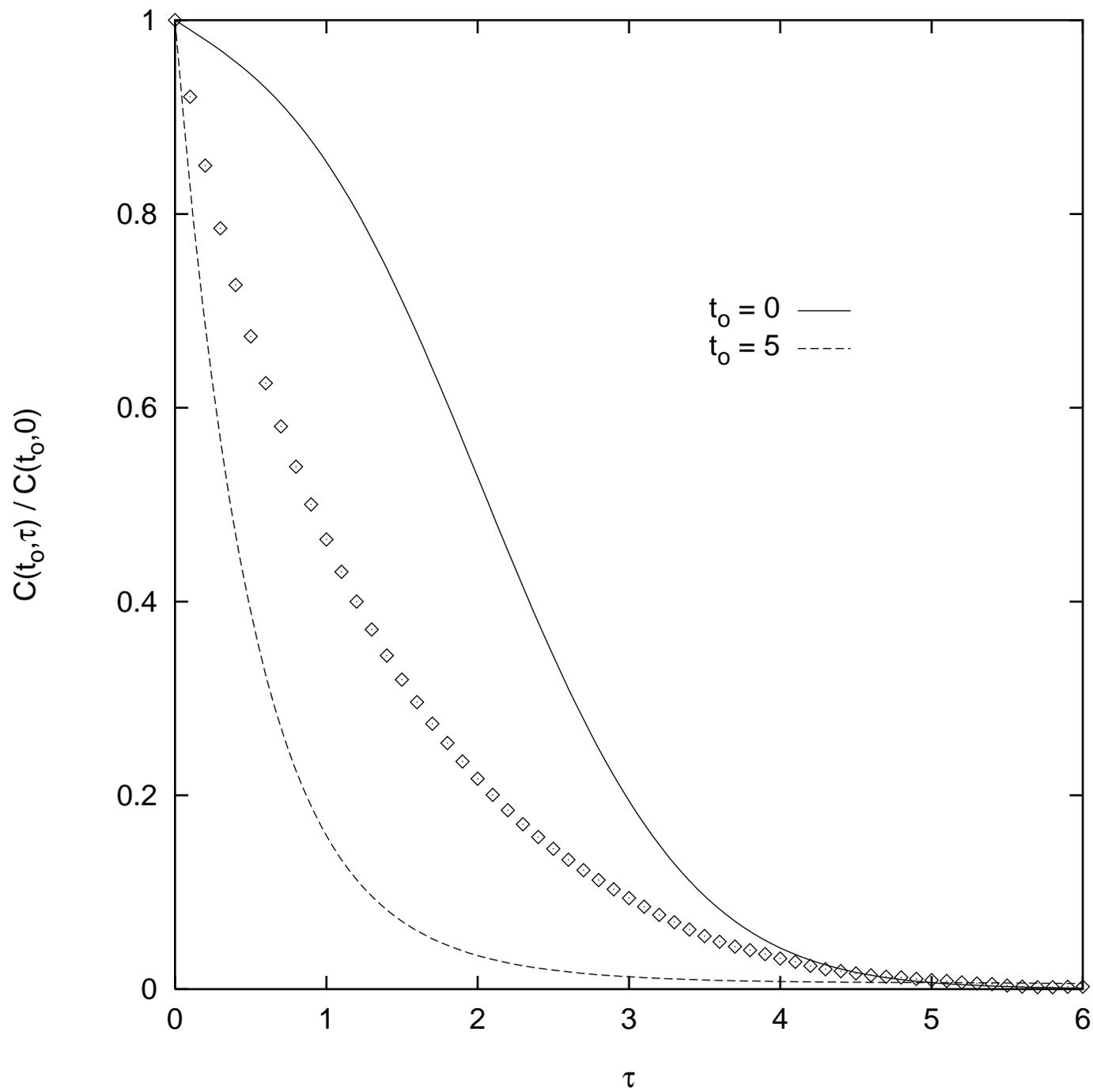
\begin{figure}
\caption{Model (\protect\ref{eq:toy1}), (\protect\ref{eq:at})
         with $a=1$, $b=-0.9$, $T=10.$ and $\Gamma=0.5$. 
        $C^{(G)}(\tau)$ (diamonds) and $C(t_o,\tau)/C(t_o,0)$ 
         as functions of $\tau$
        for $t_o=0$ (full line) and $t_o=T/2=5$ (dashed line). }
 \label{fig:toyf1}
 \end{figure}

Without the noise term, the process described by eq. (\ref{eq:toy1})
is nonchaotic. Therefore the model (\ref{eq:toy1}) 
cannot exhibit the full rich behaviour of chaotic systems, 
nevertheless it catches some of the relevant features.
It is easy to see that the characteristic decay time of 
the local correlation
\begin{eqnarray}
\label{eq:twop}  
  C(t_o, \tau)  &=& \langle q(t_o)\,q(t_o + \tau) \rangle_{t_o} \nonumber\\
           &=&\lim_{N\to\infty} \frac{1}{N}\,\sum_{k=0}^{N-1}\, 
               q(kT+t_o+\tau)\,q(kT+t_o)
\end{eqnarray}
depends on $t_o$. This can be easily computed by using 
the formal solution of (\ref{eq:toy1})
\begin{equation}
\label{eq:solu}
  q(t) = G(t)\,\left[\, q(0) + \int_{0}^{t}d\tau\, 
 G^{-1}(\tau)\,\xi(\tau)\,\right]
\end{equation}
where
\begin{equation}
\label{eq:Gt}
  G(t) = \exp\left[ -\int_{0}^{t}d\tau\, a(\tau) \right]
\end{equation}
A straightforward calculation leads to
\begin{equation}
\label{eq:cor1}
  C(t_o,\tau)  = C(t_o,0)\, G(t_o, \tau)  / G(t_o)
\end{equation}
where the equal-time correlation is
\begin{equation}
\label{eq:equalt}
 C(t_o,0) = G^{2}(t_o)\left[\, \lim_{N\to\infty}\,\frac{1}{N}\,
                          \sum_{k=0}^{N-1}\, q(kT)^2
                          + 2\Gamma\int_{0}^{t_o}d\tau\,G^{-2}(\tau)
                       \,\right]
\end{equation}
In fig.~\ref{fig:toyf1} we show $C(t_o, \tau) /C(t_o,0)$ as a function 
of $\tau$ for
\begin{equation}
\label{eq:at}
 a(t) = a + b\cos\left(\frac{2\pi}{T}\, t\right), \qquad
 a + b > 0
\end{equation}
for two different values of $t_o$, namely $t_o=0$ 
and $t_o=T/2$, with $T=10$, $a=1$, $b=-0.9$ and $\Gamma = 0.5$. 
The different behaviour is evident. By defining a 
characteristic time as the time $s$ it takes to have 
$C(t_o,s)= 0.1$, we get for this 
case $s_{0} \approx 3.475$ and $s_{T/2}\approx 1.275$. 
When starting from $t_o=T/2$ the decay is almost a factor 
$3$ faster than starting from $t_o=0$. The usual global average, 
\begin{equation}
\label{eq:twoglob}
  C^{(G)}(\tau) = \lim_{t\to\infty}\, \frac{1}{t}\,\int_{0}^{t} dt'\,
                q(t')\,q(t'+\tau)
\end{equation}
gives an average correlation function, so its characteristic decay time
is not able to distinguish different regimes. Moreover, while 
$C(t_o, \tau) / C(t_o,0)$ does not depend on the noise 
strength $\Gamma$, $C^{(G)}(\tau) / C^{(G)}(0)$ does.
In fig.~\ref{fig:toyf1} we used $\Gamma = 0.5$. 

We consider now how the system responds at time $t_o+\tau$ 
to a perturbation done at time $t_o$. This is described by 
the mean response function $R(t_o,\tau)$, which can be computed 
as follows. One takes two trajectories differing at time $t_o$ 
by a quantity $\epsilon$, i.e., $\delta q(kT+t_o) = \epsilon$ 
for any $k$,  and evolving with the {\it same} realization of 
noise. Then the local response function is
\begin{eqnarray}
\label{eq:clires}
 R(t_o, \tau)  &=& \lim_{N\to\infty} \frac{1}{N}\,\sum_{k=0}^{N-1}\, 
               \frac{\delta q(kT+t_o+\tau) }{\delta q(kT+t_o)} 
\nonumber \\
  &=& \lim_{N\to\infty} \frac{1}{N}\,\sum_{k=0}^{N-1}\, 
               \frac{\delta q(kT+t_o+ \tau) }{\epsilon}
\end{eqnarray}
where $\delta q(kT+t_o+\tau)$ is the difference 
between the two trajectories at time $t_o+ \tau$. 
Both times $t_o$ and $\tau$ run over a cycle, i.e., in the
interval $[0,T]$.

By making use of (\ref{eq:solu}) it is easy to see that 
\begin{equation}
\label{eq:clires1}
  R(t_o, \tau)  = \frac{G(t_o, \tau) }{G(t_o)}.
\end{equation}
By combining eq. (\ref{eq:cor1}) and (\ref{eq:clires1}) we have the 
fluctuations/relaxation relation \cite{Kubo85} 
\begin{equation}
\label{eq:flurel1}
  C(t_o, \tau)  = C(t_o,0)\,R(t_o, \tau) .
\end{equation}

The scenario just described remains basically valid for 
non-linear Langevin equation. Consider for example the equation
\begin{equation}
\label{eq:toy22}
\frac{d}{dt}\,q(t) = - a (t) \,q^3(t) + \xi(t)
\end{equation}
where $a (t)$ is still given by (\ref{eq:at}). 

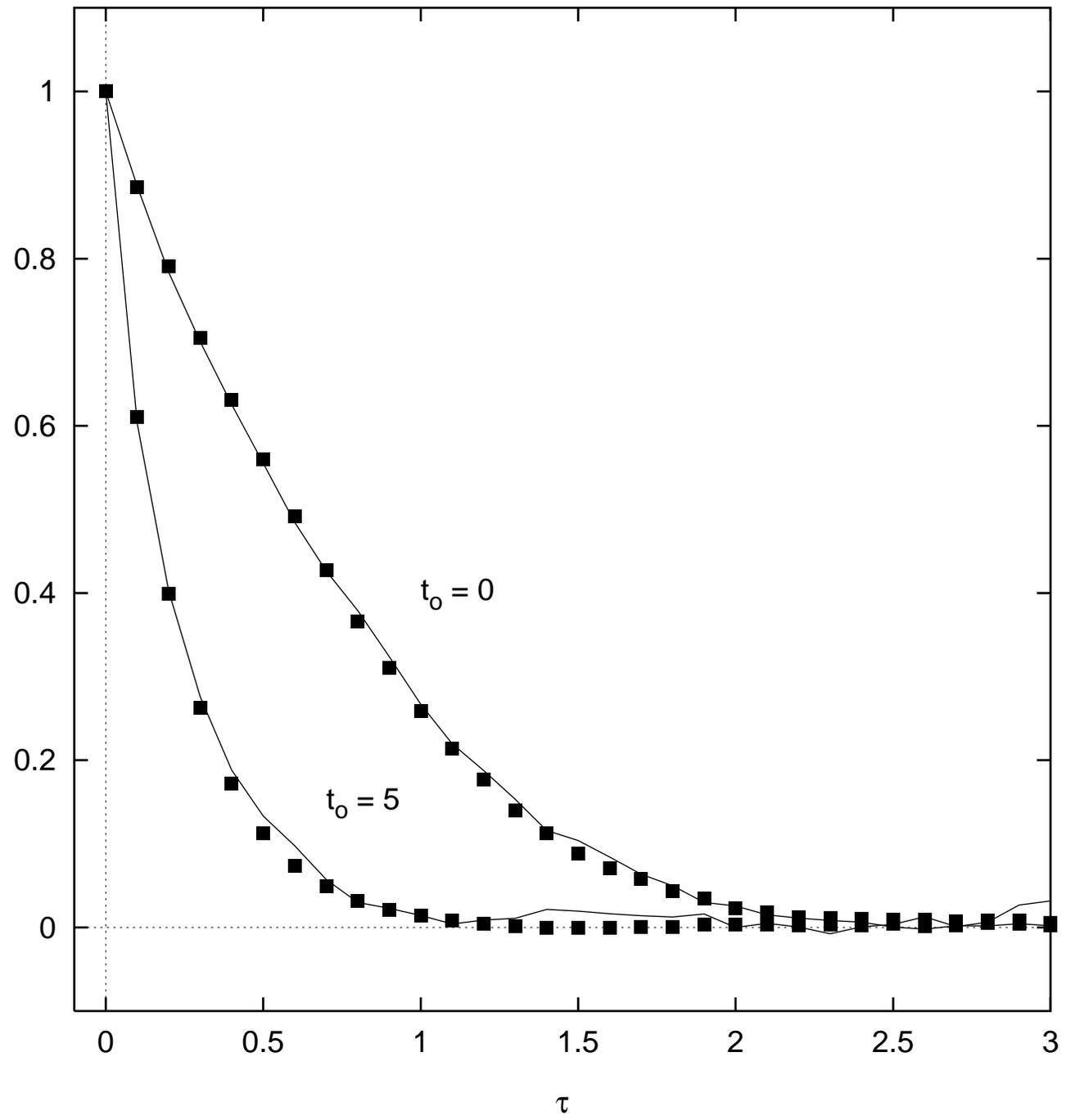
\begin{figure}
\caption{Model (\protect\ref{eq:toy22}), (\protect\ref{eq:at})
         with $a=1$, $b=-0.9$, $T=10.$ and $\Gamma=0.5$.
     $R(t_o, \tau)$ (full line) and $ \left\langle\, q(t_o, \tau) \,
     [\partial S_{t_o} (q)/\partial q(t_o)]\right\rangle _{t_o} $
    (full squares) as functions of $\tau$, for $t_o=0$ and $t_o=T/2=5$. }
 \label{fig:toyf2}
 \end{figure}

It is natural to expect that, because of the adiabatic assumption, 
for eq. (\ref{eq:toy22}) a probability distribution 
\begin{equation}
\label{eq:FDT1}
   P_t (q) \propto \exp-[S_t(q)],
\end{equation}
with $S_t(q) = q^4 a(t)/4 \Gamma$, exists at any time. 
Therefore a natural ansatz is to assume that (\ref{eq:fluri}) 
becomes: 
\begin{equation}
\label{eq:FDT}
   R(t_o, \tau)  = \left\langle\, q(t_o, \tau) \,
     \frac{\partial S_{t_o} (q)}{\partial q(t_o)} 
     \right\rangle _{t_o}
\end{equation}
In Fig.~\ref{fig:toyf2} we show $R(t_o, \tau) $ and 
$ \left\langle\, q(t_o, \tau) \,
  [\partial S_{t_o} (q)/\partial q(t_o)]\right\rangle _{t_o} $
versus $\tau$, for different $t_o$, with $a=1$, $b=-0.5$, 
$T=10$ and $\Gamma=0.5$. The two curves refer to $t_o=0$ and 
$t_o=T/2$. We see that eq. (\ref{eq:FDT})  is well obeyed and 
that the characteristic decay times are different.

 \section{Numerical simulations}
 \label{sec:numsim}

 In this section we report the results obtained for the 
 Lorenz model (\ref{eq:L3}) 
 with the Rayleigh parameter $r$ varying periodically in 
 time according to eq. (\ref{eq:L4}). We remind that 
 at the value $r=r_c=166.07$ the standard 
 Lorenz model has a transition from a regular evolution 
 (stable orbit) to a regime of intermittent chaos, and 
 the maximum Lyapunov exponent depends on $r > r_c$ through 
 a scaling law $\lambda \sim \sqrt{r-r_c}$, for 
 $r-r_c \ll 1$ \cite{Manne80}. 
 Since we are interested in chaotic systems with a non 
 negligible variation of the degree of chaoticity, we chose 
 $r_o$ close to $r_c$. Let us recall that $\lambda$ changes 
 very slowly for $r$ near $r_{cr}=24.74$, so that a periodic 
 variation of $r$ in this region is not very interesting, at 
 least if $r(t) > r_c$ for any $t$.  We take $r_o=166.6$, 
 $A=0.5$ and we consider three different periods:
 $T=1000,\, 100, \, 10$. If $T$ is very large with respect 
 to the period ($\sim O(1)$) of the unstable periodic orbit,
 the variations of $r$ can be taken as quasi-adiabatic.
 Therefore we expect that for large $T$ and for $\tau \ll T$, 
 the local effective Lyapunov exponent of the system varies 
 periodically in time with the same period of $r$,{\it i.e.},  
\begin{equation}
\label{eq:gammadr}
 \langle \gamma \rangle_{t_o} \approx \lambda [r(t_o)]
 \sim \sqrt {r(t_o)-r_c} \propto \sqrt{cos(2\pi t_o/T)}.
\end{equation}
 From fig.~\ref{fig:4}~a one sees that this approximation holds true 
 if $r(t_o)$ is not too far from $r_c$. 
 The value of the maximum Lyapunov exponent should be close 
 to the maximum Lyapunov exponent of the stationary Lorenz 
 model computed at $r_o$, {\it i.e.}, the average value 
 of $r$ on the whole period.

 When $T$ is not long enough, so that one 
 cannot speak anymore of adiabatic attractors, the 
 behavior of $\langle \gamma \rangle_{t_o}$ has 
 no relation with that of $r(t)$, as shown in fig.~\ref{fig:4}~b. 

 We discuss now the behaviour of the "local" correlation 
 functions, defined by eq. (\ref{eq:10}), and of the ``local'' 
 response functions, eq. (\ref{eq:13}). From fig.~\ref{fig:5} 
 one sees rather clearly how the ``local'' correlation functions 
 depend rather strongly from the initial time  $t_0$. In 
 particular, the dependence on $t_0$ of the mean decay time 
 is well evident.

 The correlation functions show a rapid 
 oscillating behaviour on times $\sim O(1)$, with a typical 
 period of the order of the mean circulation time near 
 the ``ghost'' of the stable orbit existing at $r=r_c$, while 
 they decay for long delays, with characteristic times that sensibly 
 depend on the degree of chaos relative to the initial instant.

 Let us stress that, because of the highly non gaussian nature of the
 system, there is not a simple direct proportionality between the 
 response $R_{ij}$ and the standard correlation $C_{ij}$, both in 
 the global and in the local version. This happens also in 
 autonomous systems \cite{Faca90} and is not a pathological behavior. 
 From very general arguments \cite{Faca90} one can show that 
 mean response and correlation function have the same qualitative 
 behavior, {\it e.g.}, the same decaying properties on large time 
 delay. However, the agreement between $R_{ij}$ and $C_{ij}$ is 
 very poor for moderate delay. This is so because in chaotic 
 systems the error bars, in the numerical computation of the 
 mean response function, increase exponentially with the delay 
 and, when $\tau$ is not very large, the mean response has to be 
 compared with a suitable correlation function 
 -- see eq. (\ref{eq:fluri}) -- which depends on the unknown 
 invariant probability distribution $P(\bbox{x})$.

 \section{Conclusions}
 We have discussed how to characterize the behaviour of a 
 chaotic dynamical system when a ``seasonal'' effect is 
 present, {\it i.e.} when there exist two well separated 
 time scales: the internal one and that one of the 
 periodic variation of a control parameter. A proper 
 characterization has been obtained by the introduction 
 of a restricted average for the relevant quantities: 
 correlation functions, response functions and Lyapunov 
 exponents. These selective averages keep into account, 
 in an explicit way, the phase of the external period 
 at which the system is observed. We stress that, in 
 the presence of a ``seasonal'' effect, the usual global 
 averaged quantities, can give only a rough information.

 \acknowledgments
 MF and AV acknowledge the financial support of the 
 Istituto Nazionale di Fisica Nucleare,
  through the {\it Iniziativa specifica FI11}.
 AC and AV acknowledge the financial support of the
  Istituto Nazionale di Fisica della Materia.

\begin{figure}
\caption{Model (\protect\ref{eq:L3}),(\protect\ref{eq:L4}) 
   with $r_o = 25.5$ and $A=4$. 
   $z$ as a function of $t/T$ for (a) $T=300$ and (b) $T=1600$. }
 \label{fig:1}
 \end{figure}

\begin{figure}
\caption{Model (\protect\ref{eq:L3}),(\protect\ref{eq:L4}) 
   with $r_o = 166,6$ and $A=0,5$. $\langle \gamma \rangle_{t_o}$ 
  as a function of $t_o$ for (a) $T=1000$ and (b) $T=10$. }
 \label{fig:4}
 \end{figure}

\begin{figure}
\caption{Model (\protect\ref{eq:L3}),(\protect\ref{eq:L4}) 
   with $r_o = 166,6$, $A=0,5$ and $T=1000$. $C_{33} (t_o,\tau)$
  as a function of $\tau$ for (a) $t_o=0$ and (b) $t_o=T/2=500$. }
 \label{fig:5}
 \end{figure}

 
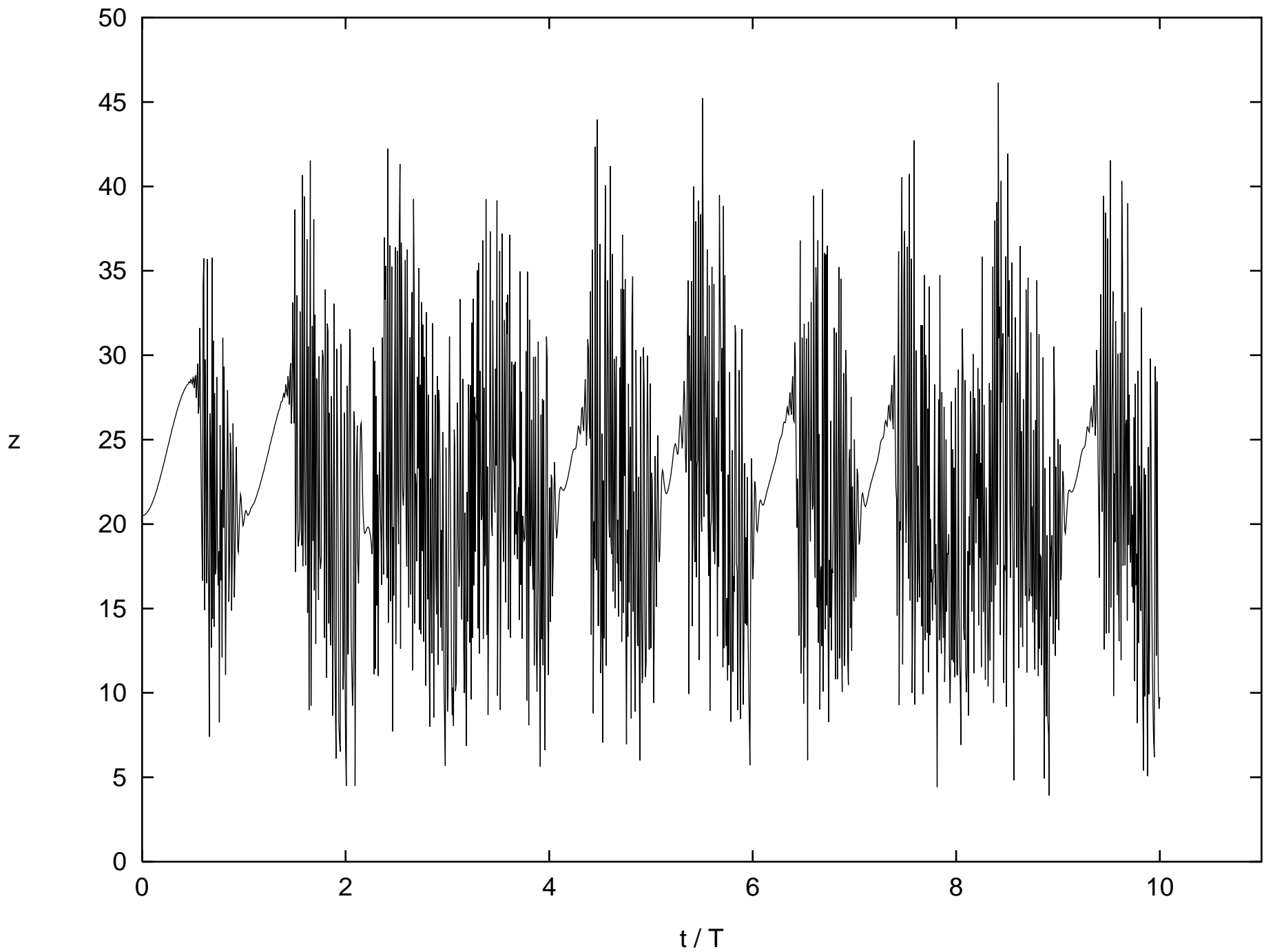
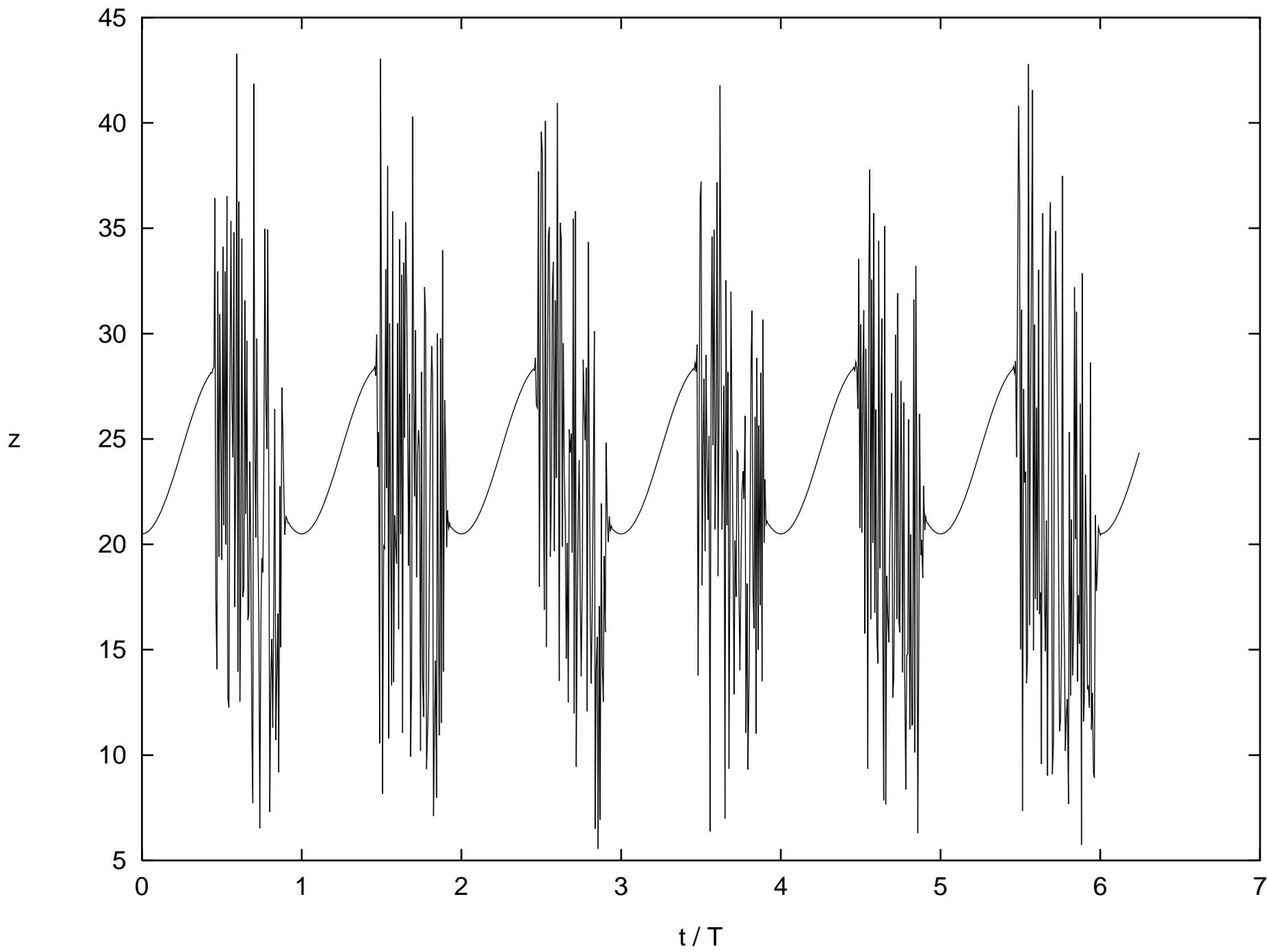
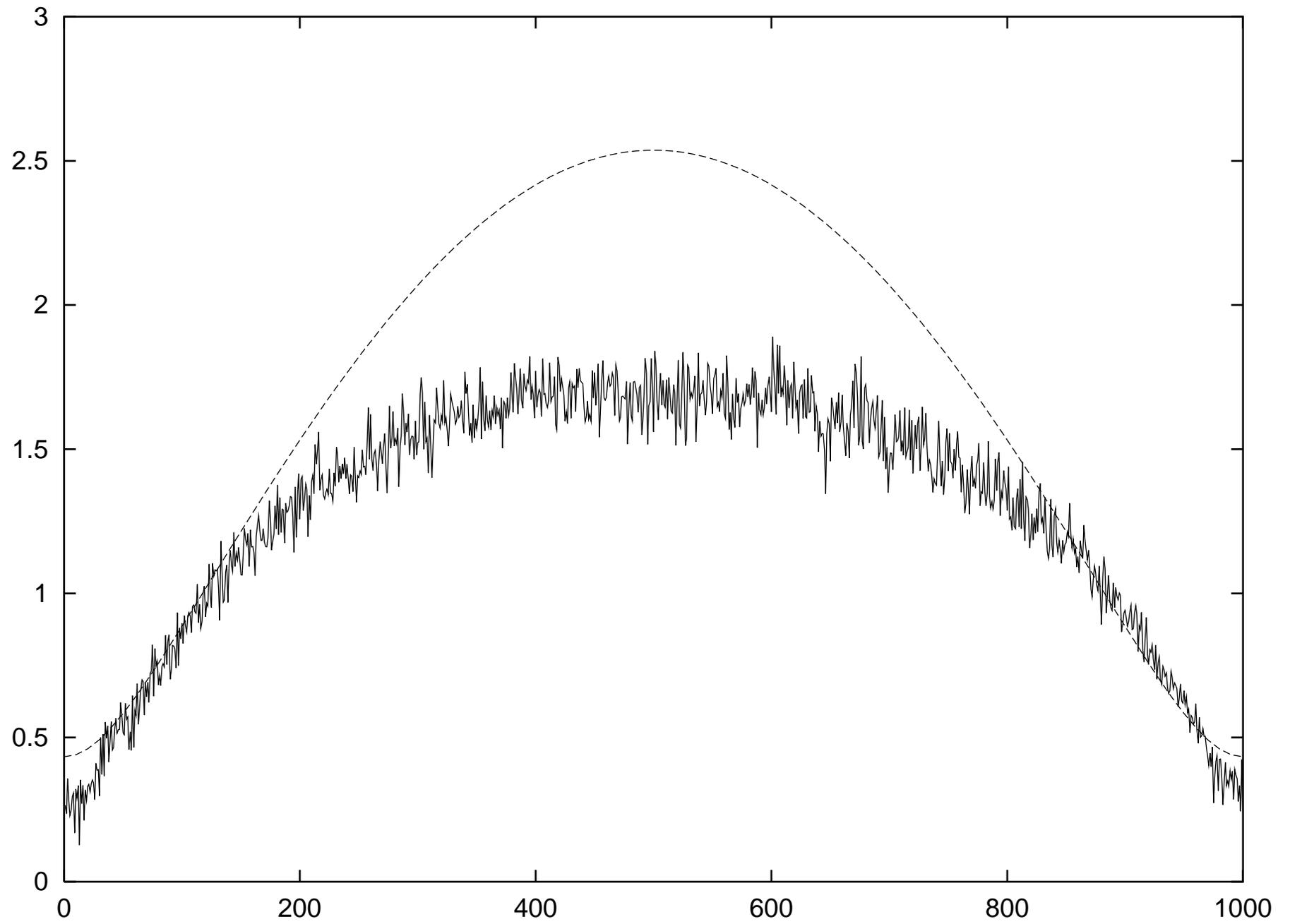
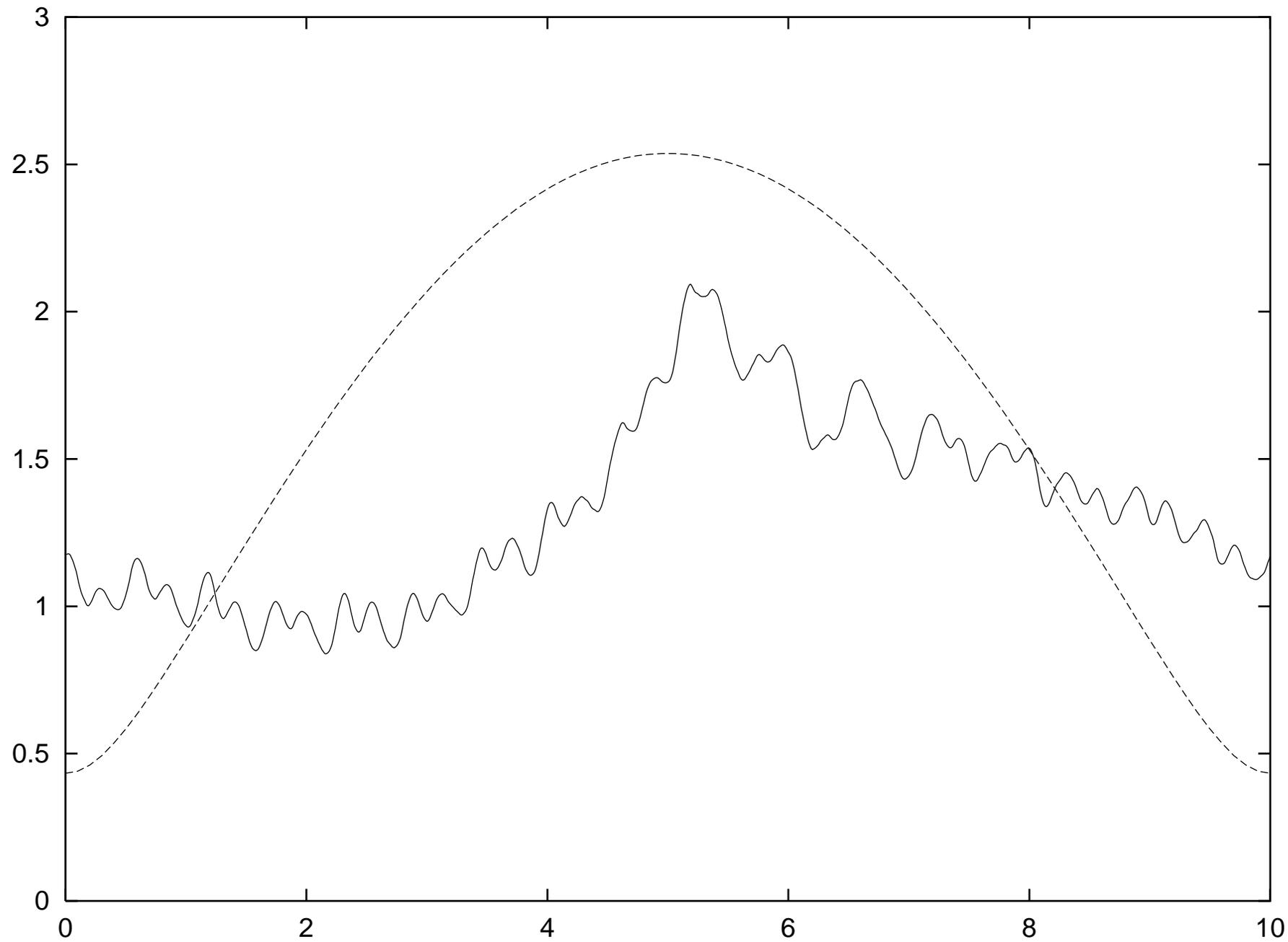
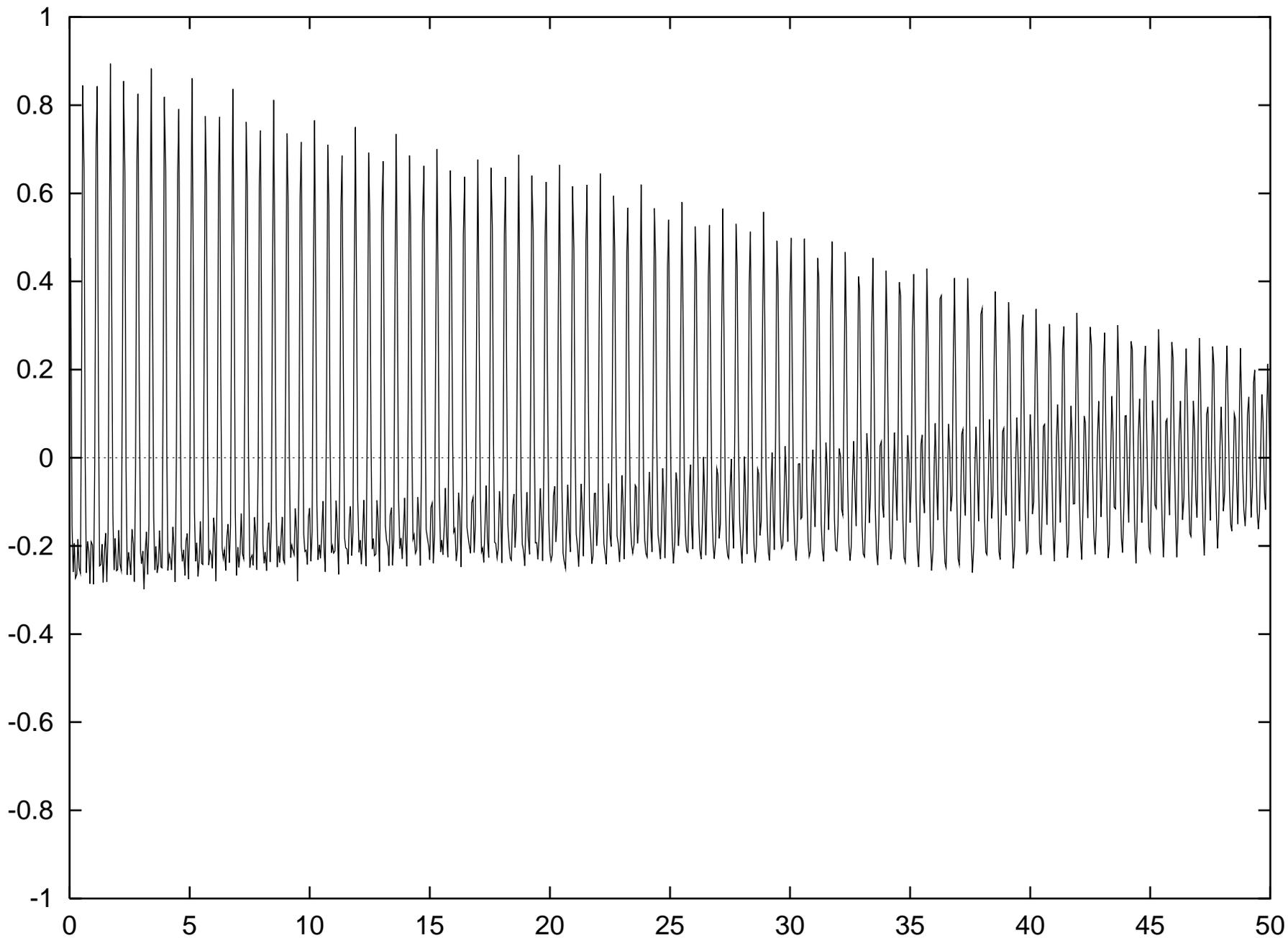
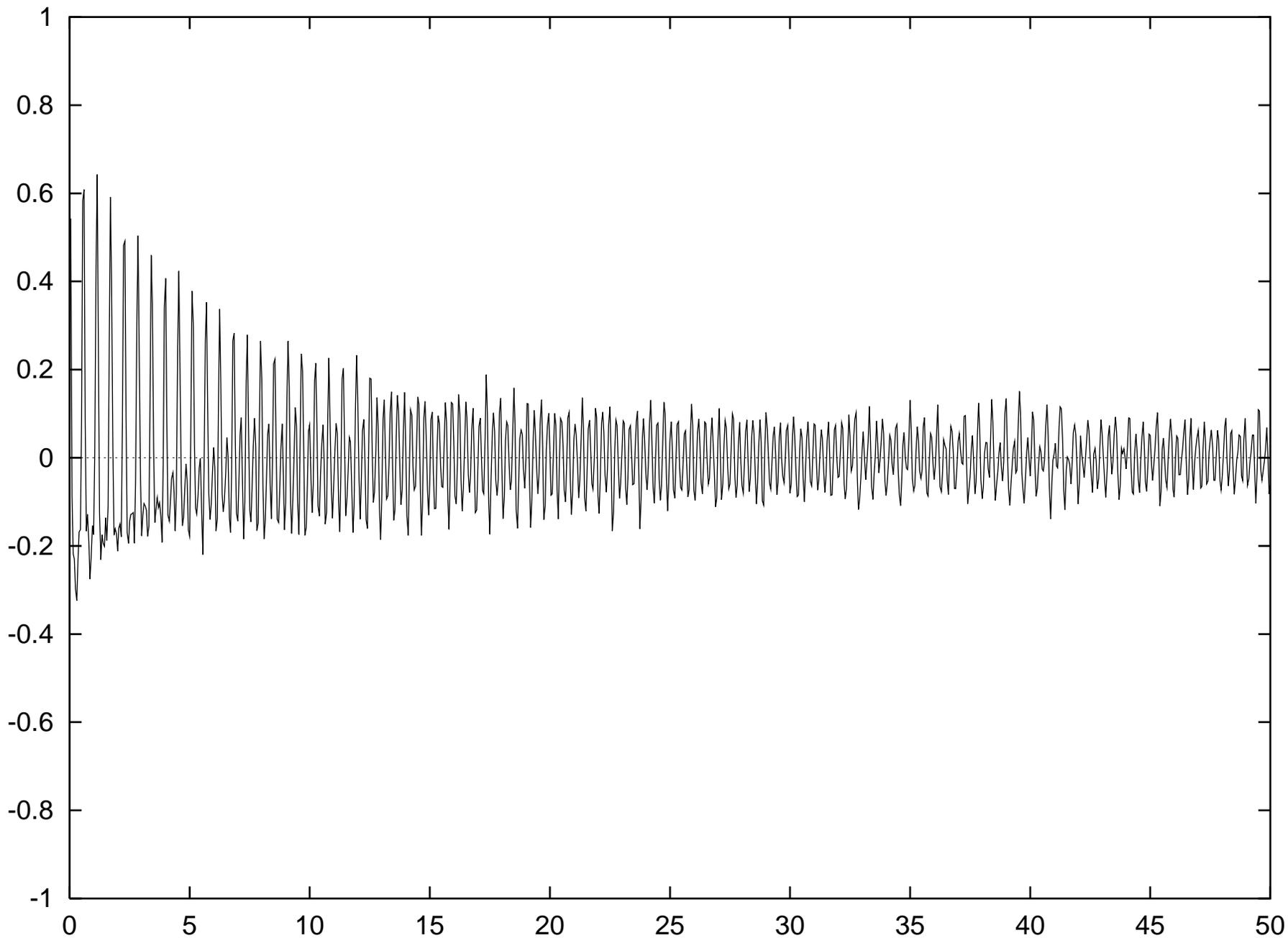
\end{document}